\documentclass[11pt]{article}
\usepackage{moriond,epsfig}

\def\be{\begin{equation}}
\def\ee{\end{equation}}
\def\bea{\begin{eqnarray}}
\def\eea{\end{eqnarray}}

\def\met{\mbox{\ensuremath{\, \slash\kern-.6emE_{T} }}}

\begin{document}
\vspace*{4cm}
\title{CHALLENGES FOR EARLY DISCOVERY IN ATLAS AND CMS}

\author{ PAUL DE JONG }

\address{NIKHEF, P.O. Box 41882 \\ 1009 DB Amsterdam, the Netherlands}

\maketitle\abstracts{
The challenges for a discovery of new physics with 1 fb$^{-1}$ of LHC data
for ATLAS and CMS are discussed. Four specific examples are chosen: a
deviation of QCD jet distributions at high $E_T$, high-mass dilepton
pairs, Higgs search in the WW decay channel, and low mass supersymmetry.
}

%%%%%%%%%%%%%%%%%%%%%%%%%%%%%%%%%%%%%%%%%%%%%%%%%%%%%%%%%%%%%%%%%%%%%%%%%%%%%%%
%
\section{Introduction}

The Large Hadron Collider (LHC) is a proton-proton collider with a center-of-mass
energy of 14 TeV, currently under construction at CERN, Geneva.
At the time of this conference~\footnote{Talk given at Moriond 2007, 
Electroweak Interactions and Unified Theories, March 10-17, 2007}, 
the LHC was still planning to have an engineering
run in the fall of 2007, in order to establish single beam operation at the
injection energy of 450 GeV, and provide first collisions at fairly
low luminosity, at 900 GeV center-of-mass energy~\footnote{However,
delays in the schedule now make this engineering run increasingly unlikely.}.
The first collisions at $\sqrt{s} =$ 14 TeV could occur in spring 2008, and 
with a steadily
increasing luminosity during the next 26 weeks of proton-proton running,
the ATLAS and CMS experiments might collect an integrated luminosity close 
to 1 fb$^{-1}$
by the end of 2008. In order to be able to present first results at
Moriond 2009, ATLAS and CMS face a number of challenges, some of
which will be discussed in this paper. Hereby we will focus on four
examples of possible signs of new physics in first data.

\begin{figure}[htbp]
  \begin{center}
    \psfig{figure=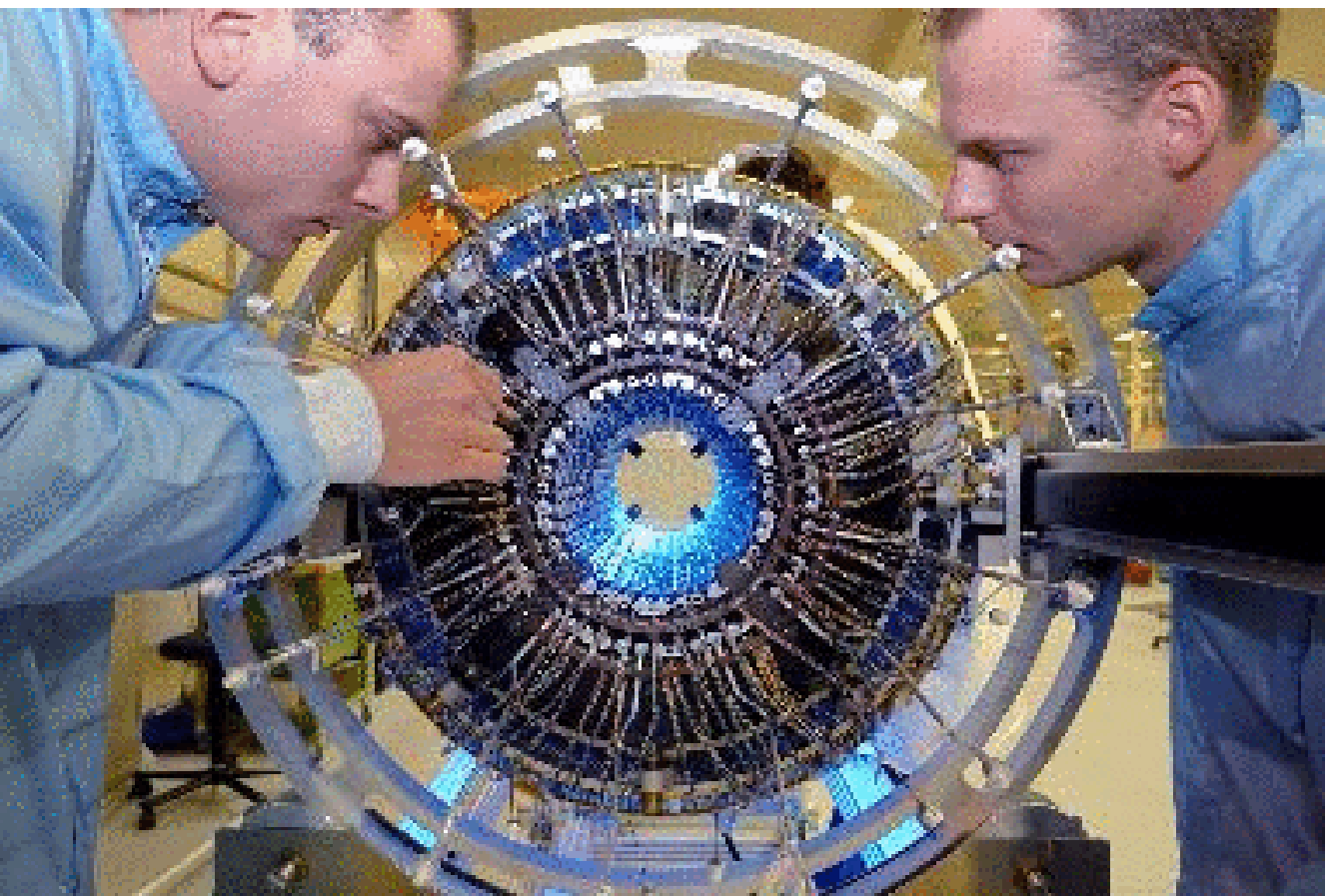,height=4cm} \hspace*{1cm}
    \psfig{figure=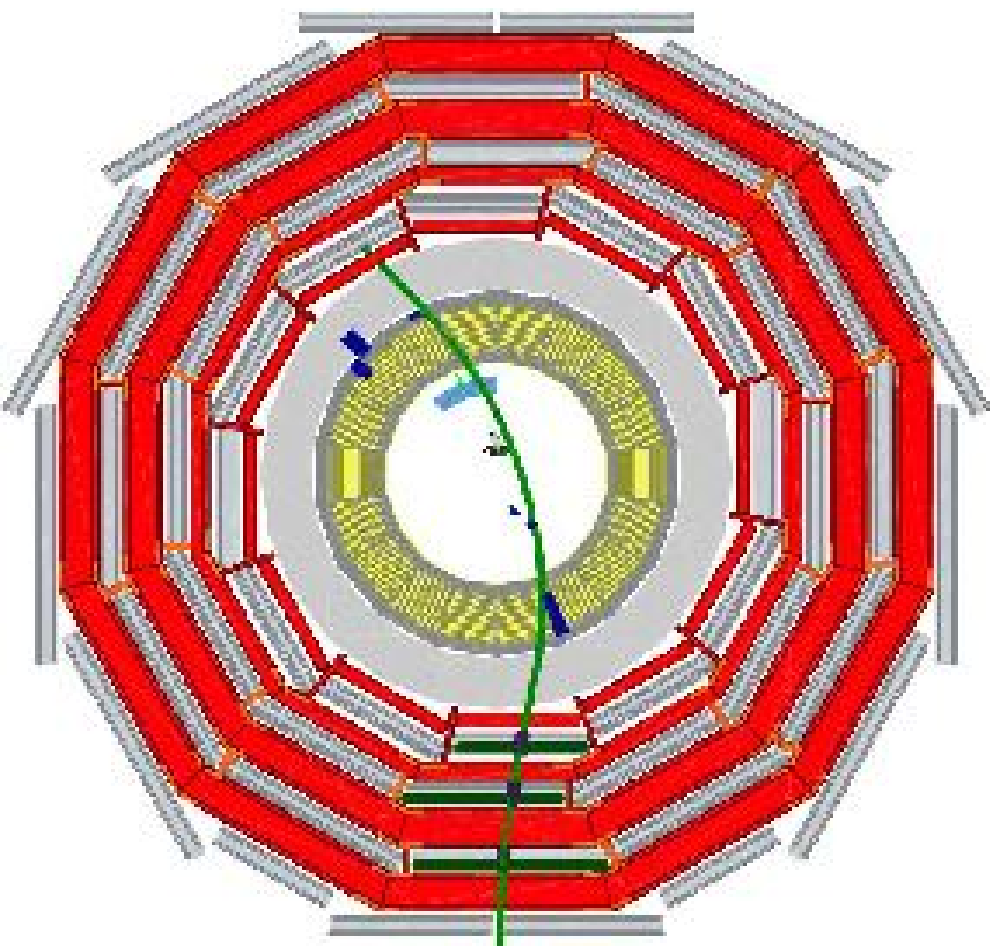,height=4cm}
  \end{center}
  \caption{Left: construction of the ATLAS pixel detector.
  Right: a muon, from a cosmic ray shower, traversing a part of CMS.
  Data taken in a commissioning run end of 2006 with CMS in the surface hall.
  \label{fig:detectors}}
\end{figure}

The first challenges that ATLAS and CMS face are still severe: the completion
of subdetector construction, installation and commissioning, establishing
reliable detector operation, and getting the trigger, data acquisition and
detector calibration infrastructure
to work as designed. The data is to be distributed and analyzed on the
Grid, and this also needs to be commissioned. ATLAS is currently undergoing
such a (Monte Carlo) data generation and analysis challenge, 
% under the terms
% of ``Computing System Commissioning'' and ``Calibration Data Challenge'', 
which will
also lead to performance estimates that update the physics TDR (technical
design report) from 1999~\cite{atlastdr}. CMS has produced an updated
physics TDR in the summer of 2006~\cite{cmstdr}.

At any energy, soft (low $p_T$) hadronic interactions, or ``minimum bias''
events, will be most common. Various
Monte Carlo generators differ significantly in their predictions of the
cross section and the charged particle multiplicity at $\sqrt{s} =$ 14 TeV.
Since these minimum bias events set the background for trigger and reconstruction,
measuring their properties will be first priority for ATLAS and CMS.
Furthermore, these events are useful for tracking studies and calorimeter 
intercalibrations.

% If the engineering run at $\sqrt{s} = 900$ GeV indeed takes place, the
% energy and luminosity will be insufficient to produce large numbers of
% W and Z bosons for detector calibration. There will be many ``minimum bias''
% events: low $p_T$ hadronic interactions, with sufficient tracks to
% perform tracking studies or calorimeter intercalibrations. Also at
% $\sqrt{s} = 14$ TeV, these minimum bias events will be most common. Various
% Monte Carlo generators differ significantly in their prediction of the
% cross section and the charged particle multiplicity at that energy.
% Since these minimum bias events set the background for trigger and reconstruction,
% measuring their properties will be first priority for ATLAS and CMS.

%%%%%%%%%%%%%%%%%%%%%%%%%%%%%%%%%%%%%%%%%%%%%%%%%%%%%%%%%%%%%%%%%%%%%%%%%%%%%%%
%
\section{QCD (di)jets at high $E_T$}

\begin{figure}[htbp]
  \begin{center}
    \psfig{figure=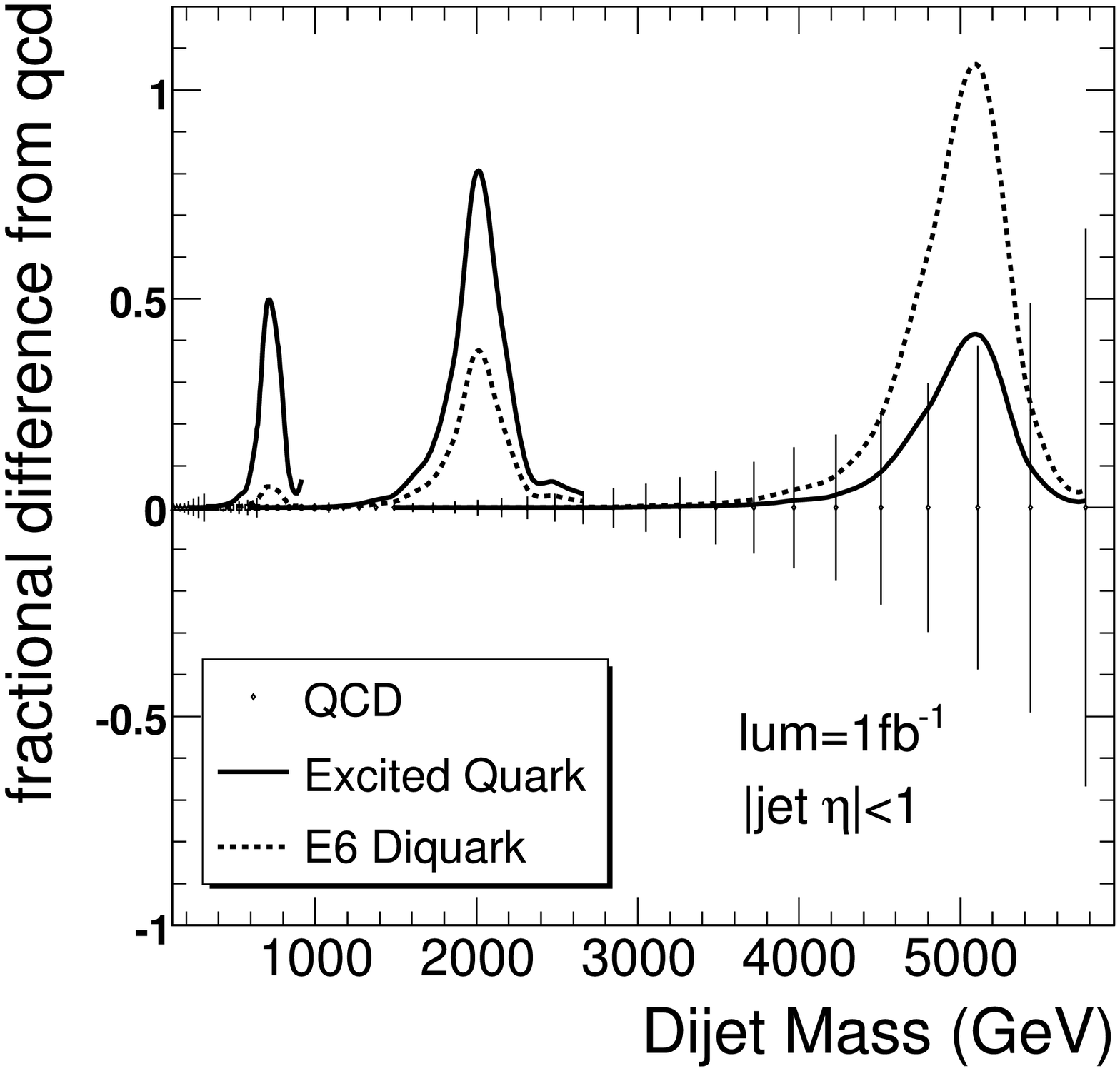,width=0.3\textwidth}
    \psfig{figure=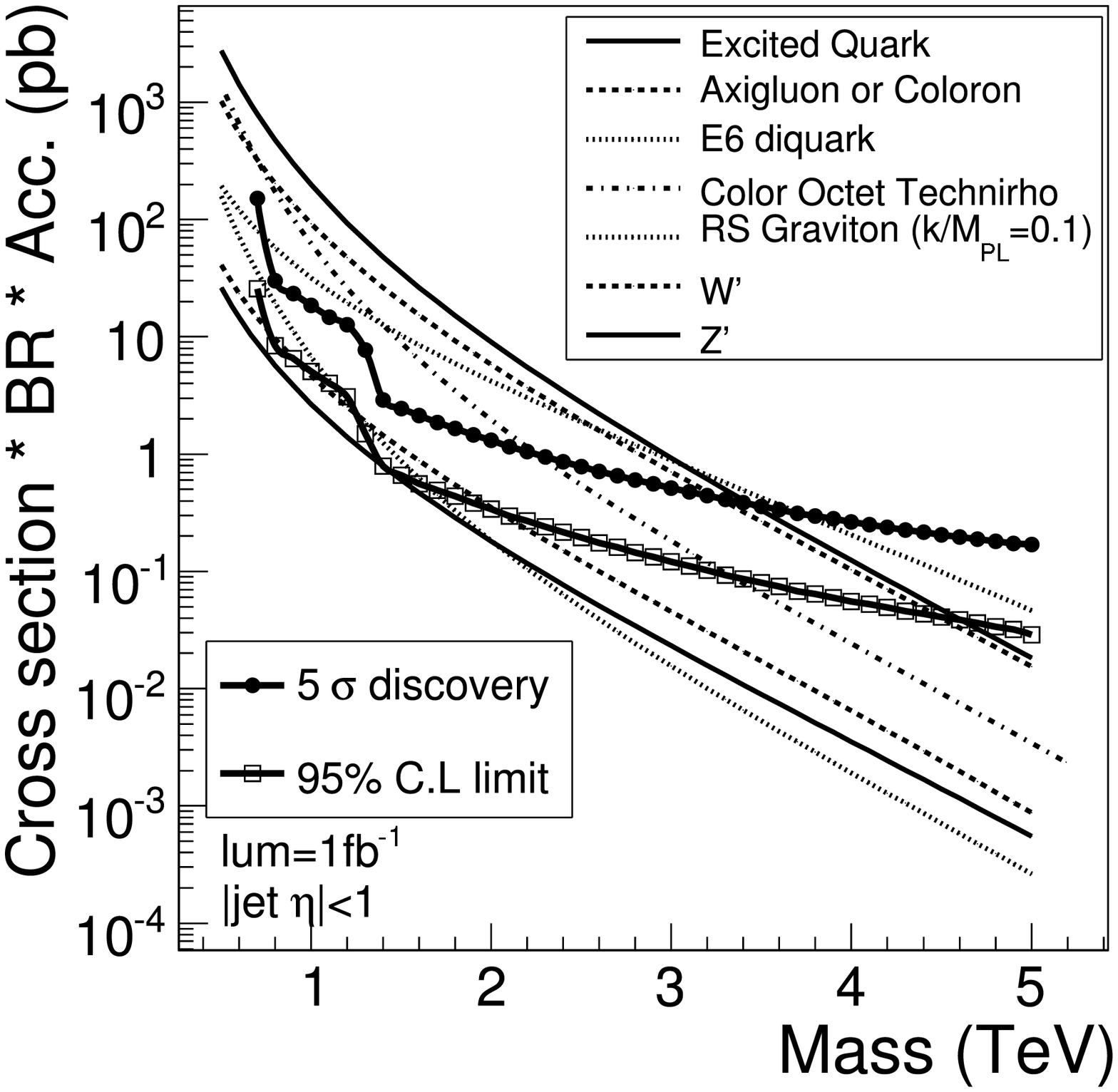,width=0.3\textwidth}
    \psfig{figure=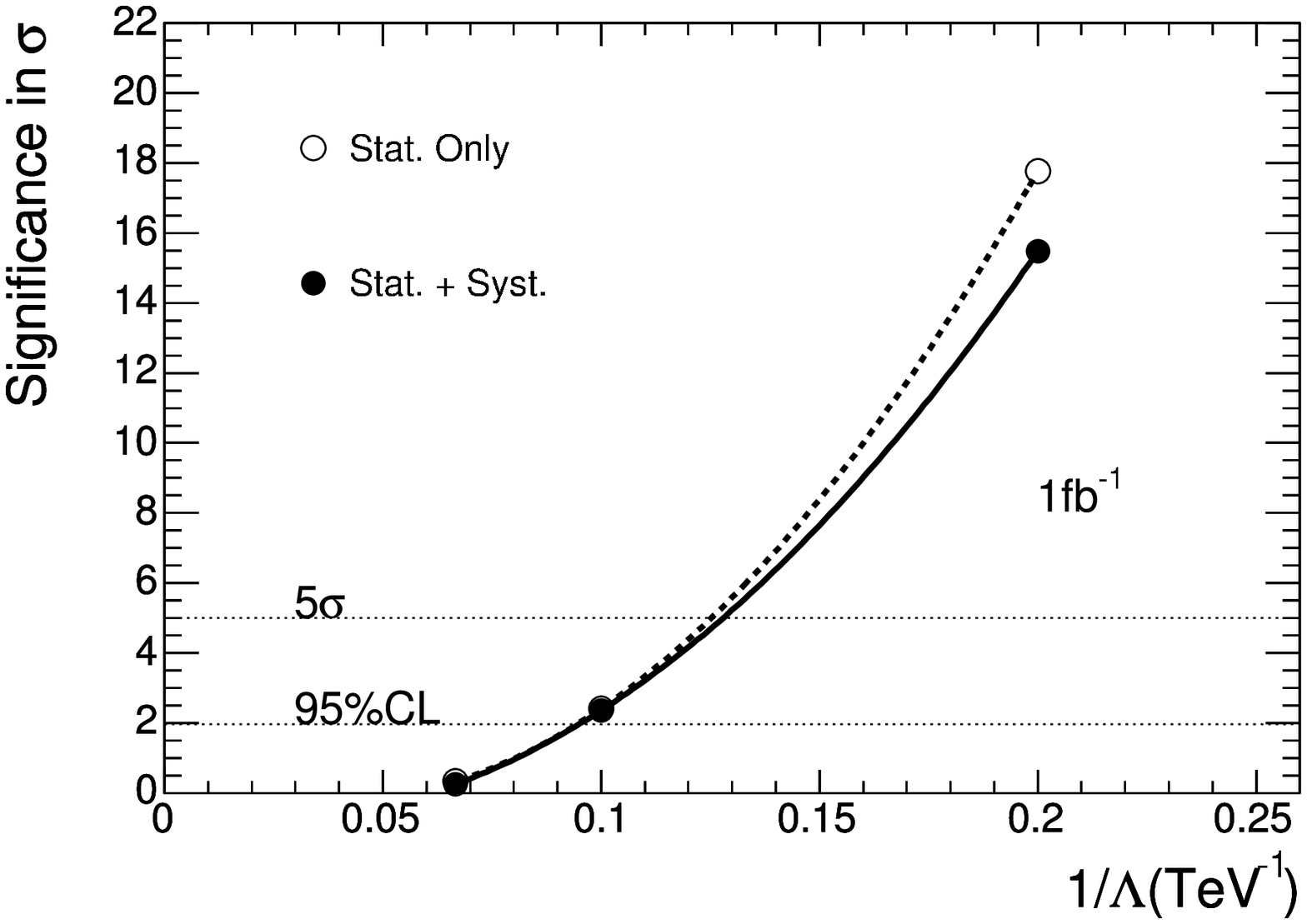,width=0.3\textwidth,height=4cm}
  \end{center}
  \caption{Left: deviations from the QCD prediction of the dijet mass
           distribution, for two hypothetical models of new physics, each
           at three different masses. The vertical bars represent the
           statistical error with 1 fb$^{-1}$ of data. Center: exclusion
           and discovery limits on the cross section for various new particles,
           as a function of the particle mass. Right: significance 
           for an exclusion or a discovery of contact interactions,
           for 1 fb$^{-1}$ of data, as a function of the (inverse of the)
           contact interactions scale. Plots from CMS.
  \label{fig:excited}}
\end{figure}

The large center-of-mass energy of the LHC gives access to
a kinematic region of QCD jet production that has never been probed before. With
1 fb$^{-1}$ of data, jets of 3-3.5 TeV transverse energy ($E_T$), and di-jet masses
of up to 6 TeV are accessible. New physics, in the form of quark substructure
and excited quarks, contact interactions, or resonances, could appear,
as shown in Figure~\ref{fig:excited}. Two
main uncertainties must be tackled before any excess can be interpreted
as new physics: parton density function (pdf) uncertainties, and the jet
energy scale.

\begin{figure}[htbp]
  \begin{center}
    \psfig{figure=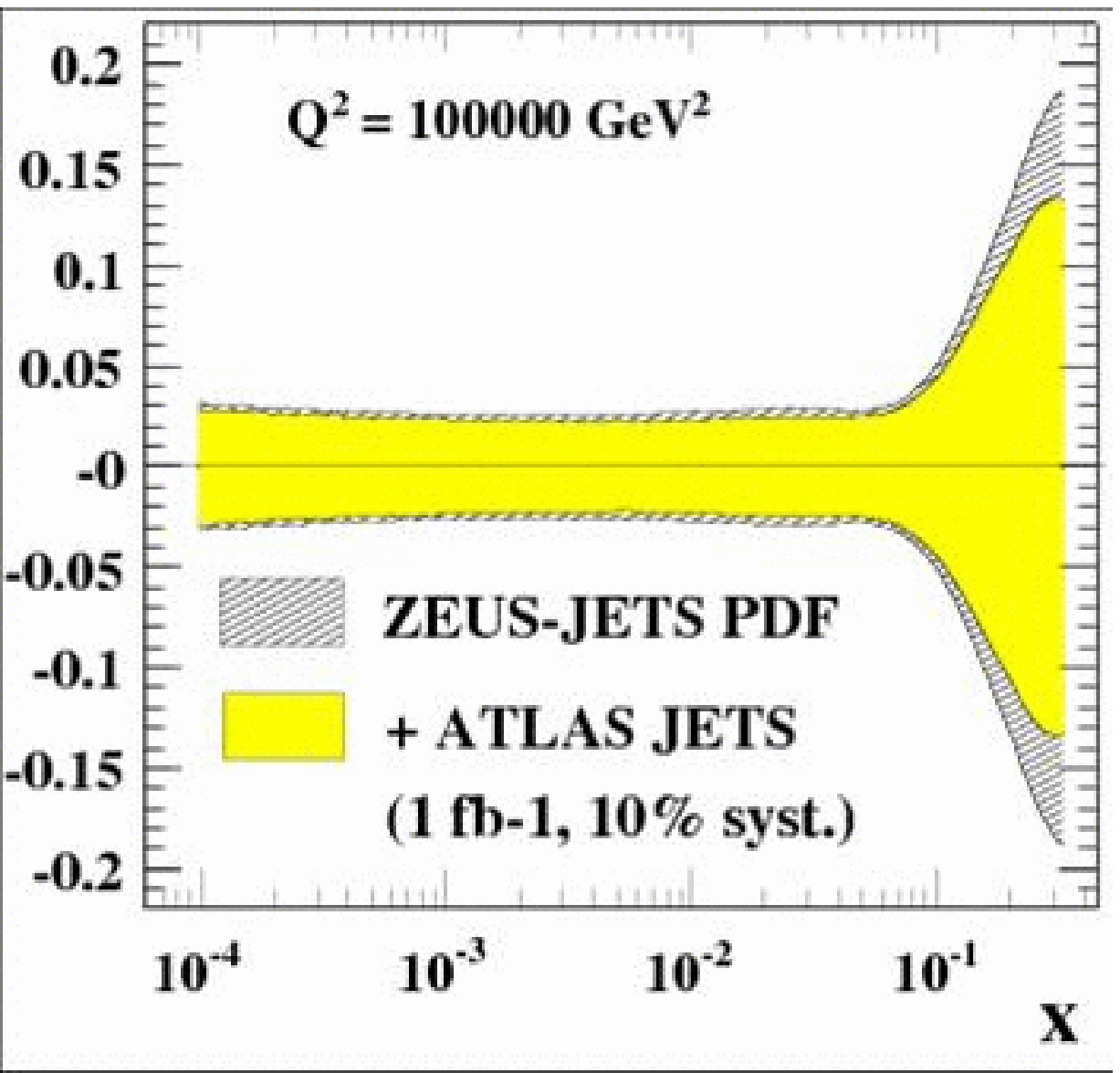,width=0.3\textwidth}
    \psfig{figure=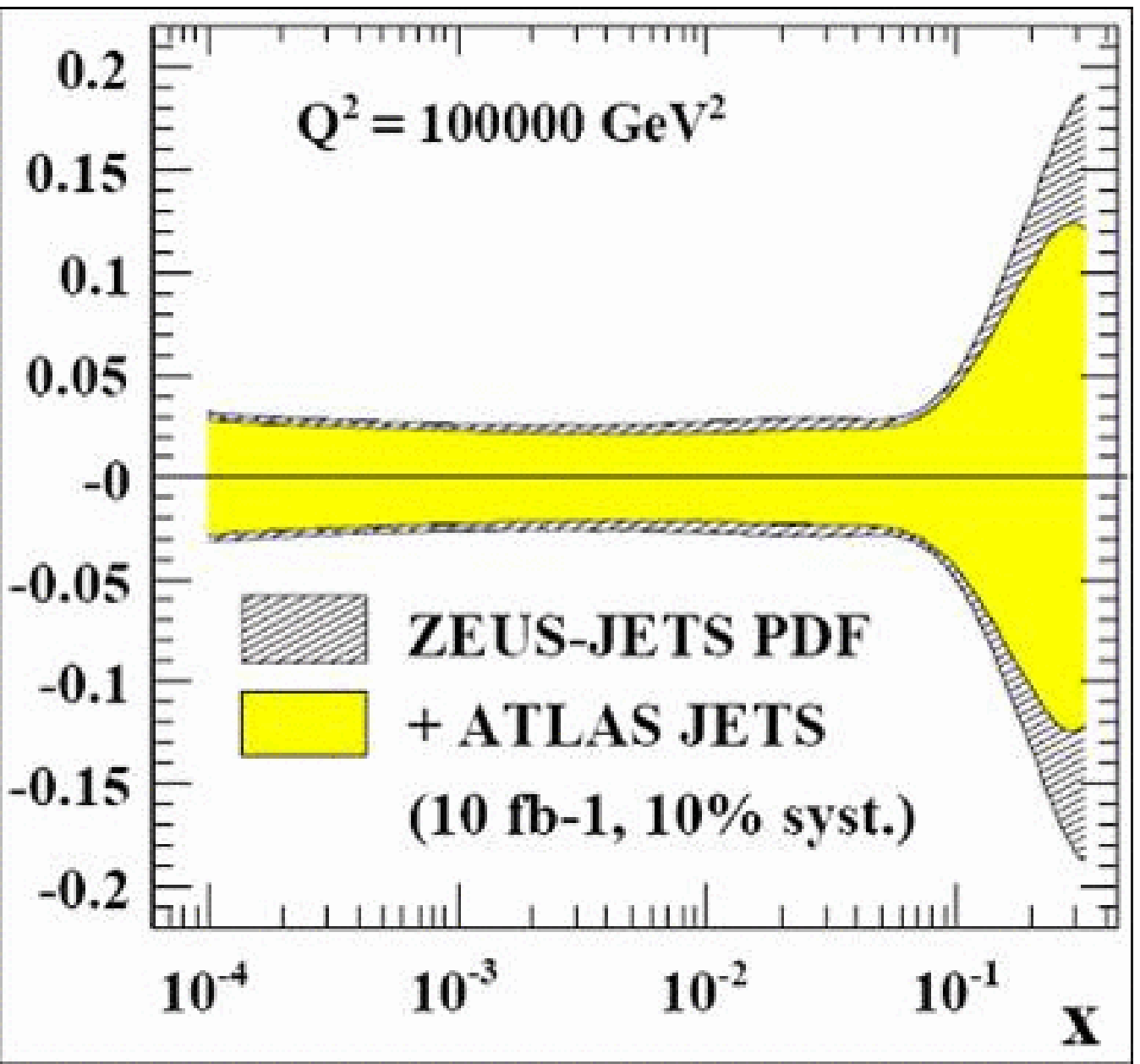,width=0.3\textwidth}
    \psfig{figure=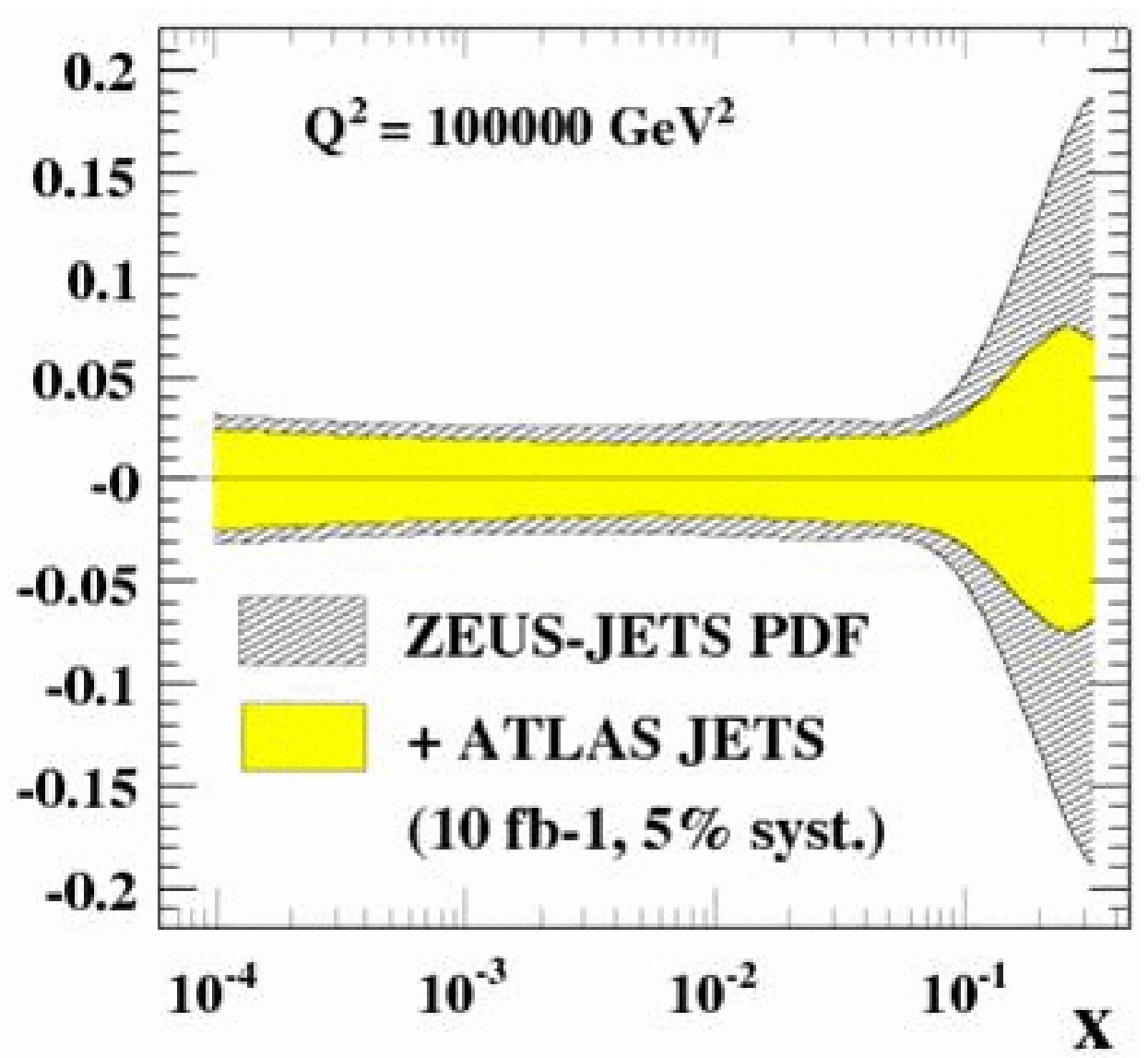,width=0.3\textwidth}
  \end{center}
  \caption{Uncertainty on the gluon distribution function as a function
  of $x$, for $Q^2 = 100000$ GeV$^2$, at this moment (ZEUS-JETS PDF), and
  including 1 fb$^{-1}$ (left) or 10 fb$^{-1}$ (center) of ATLAS jets data
  with a 10\% systematic uncertainty, and including 10 fb$^{-1}$ (right) of
  ATLAS data with a 5\% systematic uncertainty.
  \label{fig:jetspdf}}
\end{figure}

The pdf uncertainties arise mainly from the uncertainties on the gluon distribution
function at high $x$. The high $E_T$ jet data from ATLAS and CMS can in fact
be used to constrain these uncertainties. In order not to sweep new physics
under the rug in such a procedure, it is important to fit also complementary
processes, like photon plus jet production, in pdf fits, and to measure
jet production over a large kinematic range: new physics is often more central,
whereas pdf effects show up over all phase space. ATLAS has shown that with
1 fb$^{-1}$ of data, already an improvement from the current situation can be 
obtained, as shown in Figure~\ref{fig:jetspdf}. Beyond that, however, 
the systematic errors on the data must be tackled: reducing these errors help
more than adding more luminosity. These systematic uncertainties concern mainly the 
jet energy scale.

The jet energy scale will initially be known only from test beam, cosmics
and calibration systems, to not better than 5-10\%. With first data, a data-driven
jet energy scale determination program must be started immediately. This is
a major effort. As a comparison: it has taken the D0 collaboration five years
of continuous effort by a large group of people to reach now 2\% uncertainty
on the energy scale of jets between 30 and 200 GeV. ATLAS and CMS aim for
2-3\% after one year, using photon plus jets, Z plus jets, and top-quark
pair events. The latter uses the W mass constraint on the two light-quark jets,
the former two need a calibration of the electromagnetic energy scale
first. Jets from b-quarks need a different calibration than light-quark jets.

Concerning new physics, CMS estimates a discovery potential with 1 fb$^{-1}$ 
of data for excited quarks up to masses of 3.4 TeV, for diquarks in an $E_6$ 
model up to 3.7 TeV, and for the scale of contact interactions up to 7.7 TeV. 
The sensitivity of ATLAS is expected to be similar.

%%%%%%%%%%%%%%%%%%%%%%%%%%%%%%%%%%%%%%%%%%%%%%%%%%%%%%%%%%%%%%%%%%%%%%%%%%%%%%%
%
\section{High mass lepton pairs}

New physics may well show up in high-mass lepton pairs. Resonances such as
Z', gravitons in the Randall-Sundrum model, or Kaluza-Klein excitations in
models with universal extra dimensions may be seen in the dilepton mass 
distribution, on top of a Drell-Yan continuum background
that falls rapidly with increasing mass. Large extra dimensions of the ADD-model
type will not show up as resonances, but as deviations from the mass and
angular spectrum of Drell-Yan lepton pairs.

Reconstruction of electrons and muons requires a well-calibrated and
aligned detector. The alignment of the CMS and ATLAS inner detectors will
be a significant task. CMS has some 20000 silicon modules, or a total of 120000
alignment parameters to be determined; ATLAS has about 6000 modules.
Initially, alignment will come from survey measurements, dedicated hardware
systems, and cosmic-ray muon data. The ATLAS silicon strip detector includes
a system of frequency scanning interferometers, that measure the movements
of whole structures (barrels, discs) very precisely over timescales of hours;
the CMS inner detector has a laser calibration system. The ATLAS muon
spectrometer has an extensive set of laser alignment systems.
Eventually, tracks from the data will provide alignment information, but
it will be a challenge to do this for all modules, with little residual
systematics.

\begin{figure}[htbp]
  \begin{center}
    \psfig{figure=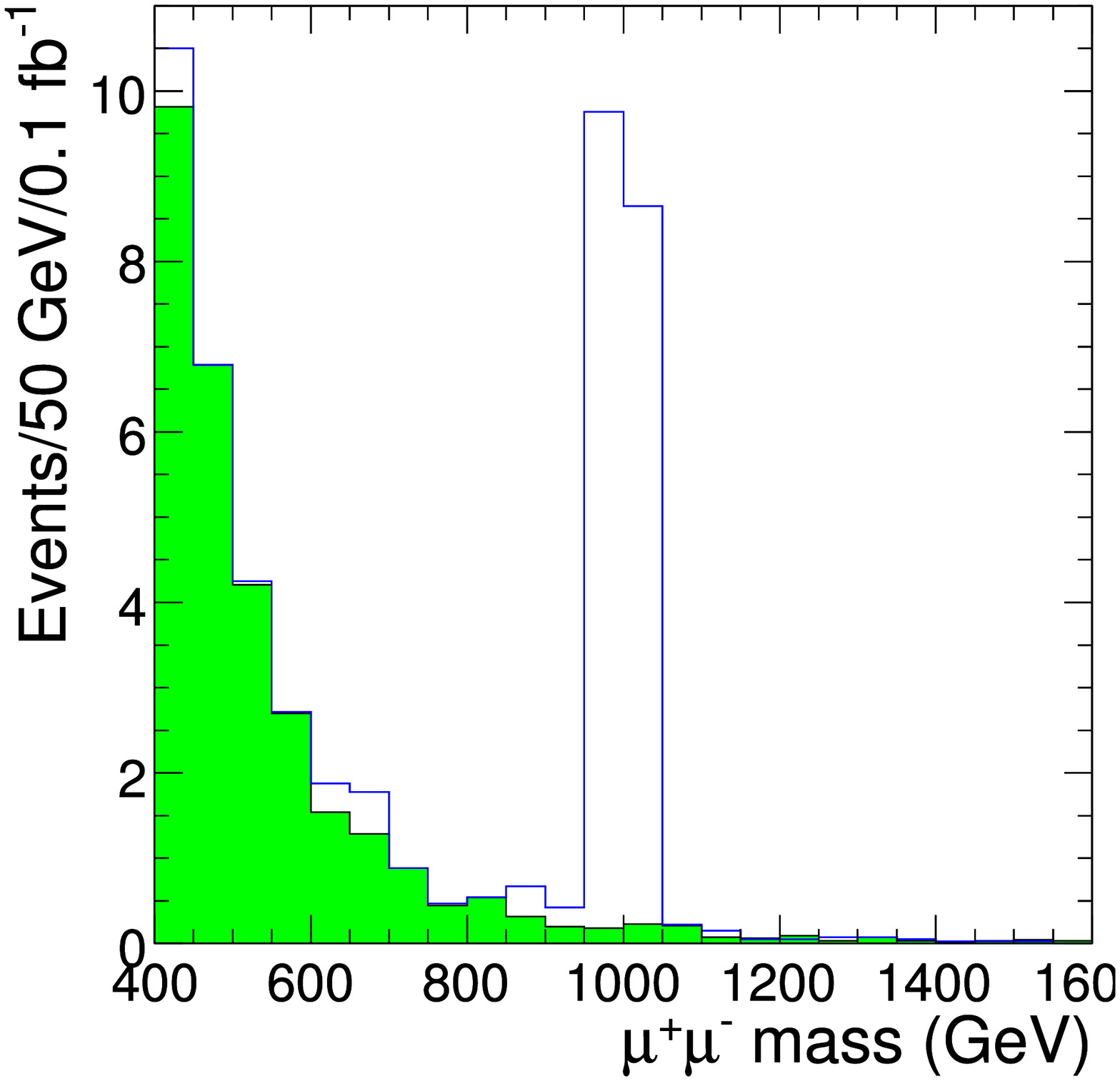,height=4.2cm} \hspace*{2mm}
    \psfig{figure=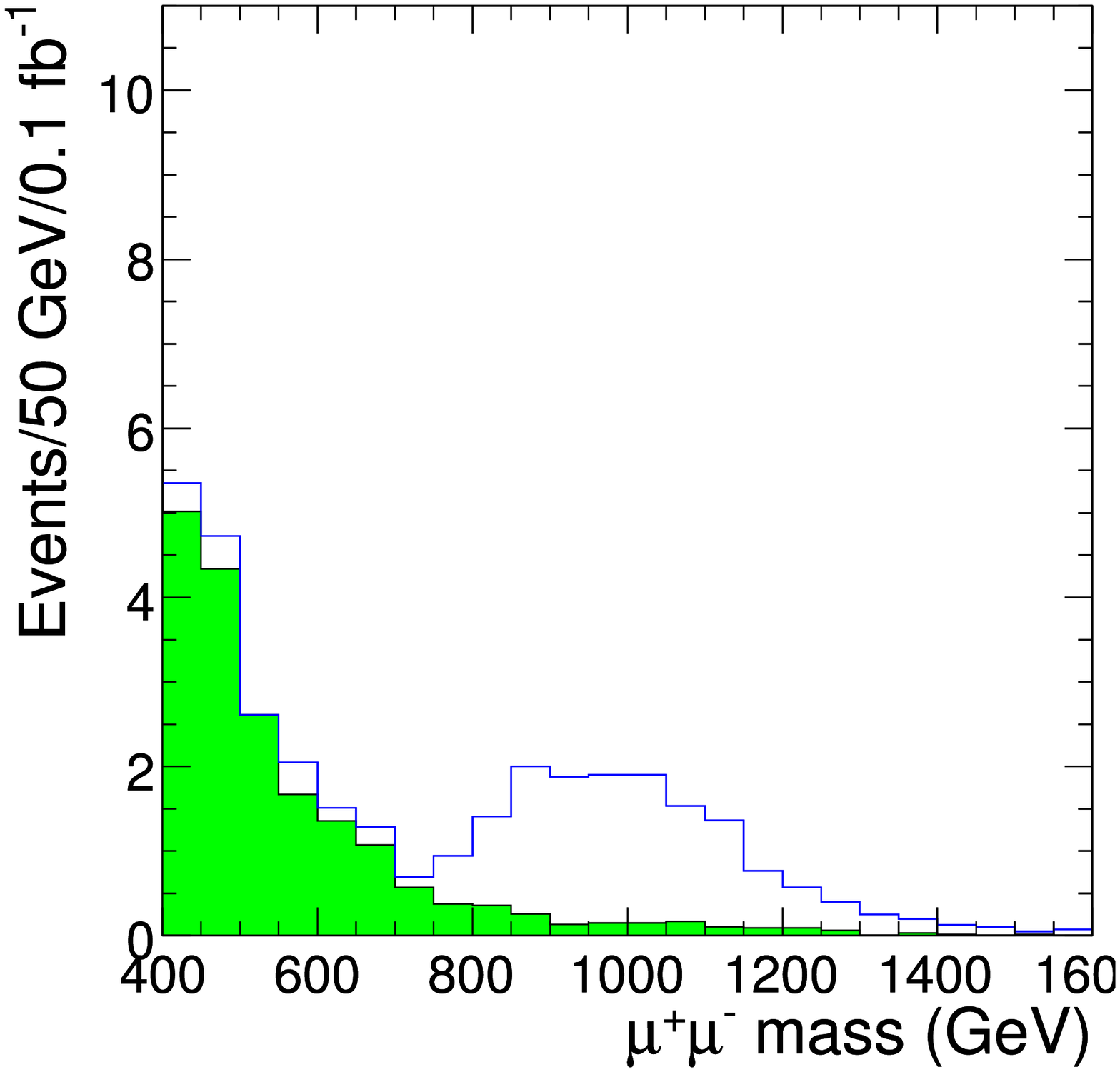,height=4.2cm} \hspace*{2mm}
    \psfig{figure=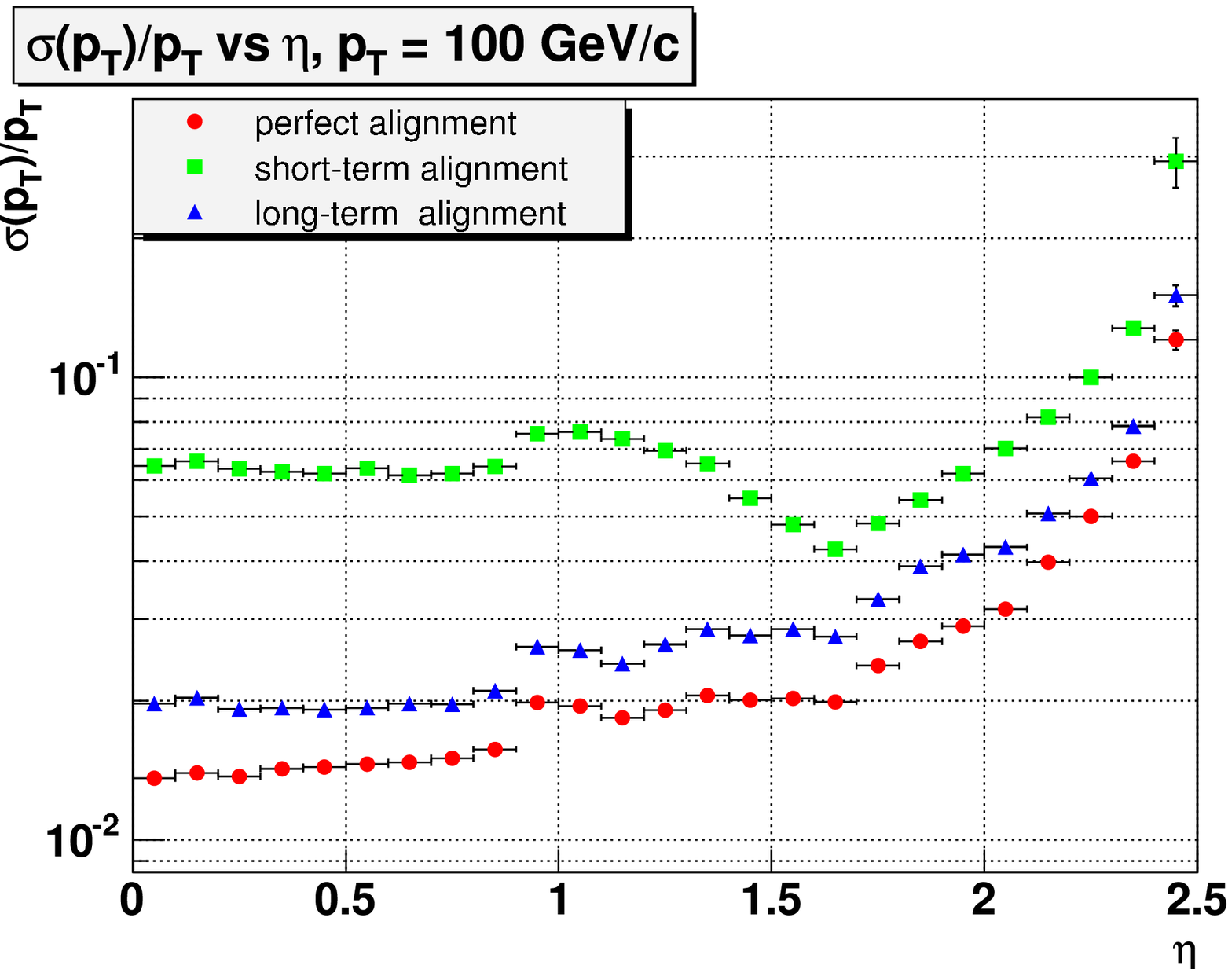,height=4.2cm}
  \end{center}
  \caption{Muon-pair mass distribution from the decay of a 1 TeV Z',
  with ideal alignment (left), or alignment as expected at LHC start-up
  (center) (also denoted short term alignment) in CMS. Right: track
  $p_T$ resolution for a $p_T = 100$ GeV track in CMS, as a function of
  pseudorapidity $\eta$, for ideal, short-term and long-term (more than
  a few fb$^{-1}$ of data) alignment.
  \label{fig:ptresolution}}
\end{figure}

The momentum scale of muons is very much determined by the alignment
accuracy of the inner detector and the muon system. It is calibrated with
muon pairs from decay of resonances such as the Z boson, and $J/\psi$ and
$\Upsilon$. Three days of data taking at $10^{33}$ cm$^{-2}$ s$^{-1}$ will
provide a sample of more than $10^5$ Z $\rightarrow \mu^+ \mu^-$ events,
and eventually the muon momentum scale can be determined to better than 0.1\%.

Electrons are measured both in the tracking system and in the electromagnetic
calorimeter (ECAL). The CMS ECAL consists of lead tungstate crystals, and
will in first instance be intercalibrated to a 0.4 to 2.0\% uniformity
with single electrons and minimum bias events. The ATLAS ECAL is a lead and
liquid argon calorimeter; the goal is a 0.4 to 1.0\% uniformity. Both
detectors will derive the overall energy scale from Z $\rightarrow e^+ e^-$
events to better than 0.1\%.

It should be noted that the energy scale is affected by many effects
(magnetic field uniformity, material in the detector, response,
alignment etc), and that all these need to be disentangled. After all,
the energy scale needs to be extrapolated from the Z peak to
high $p_T$, where the new physics is expected. A careful Monte Carlo
modeling is crucial.

Trigger and reconstruction efficiencies must be obtained from the data
itself, and it is important to include redundant and 
as-little-bias-as-possible triggers in the trigger menu. 
Also redundant object reconstruction
methods are needed: e.g. muons must be reconstructed in the inner detector,
in the calorimeter, and in the muon system, so that efficiencies and
fake rates can be determined.

On the theoretical side, there are still uncertainties on the Standard
Model prediction, originating from missing higher orders in the calculation,
scale variations, and pdf uncertainties. ATLAS and CMS will try to select
control samples from data to measure the Standard Model background, but one
should realize that new physics will probably show up in ``untypical''
corners of phase space. Therefore, good Monte Carlo predictions are still
important, and some NLO calculations are still needed~\cite{huston}.

\begin{figure}[htbp]
  \begin{center}
    \psfig{figure=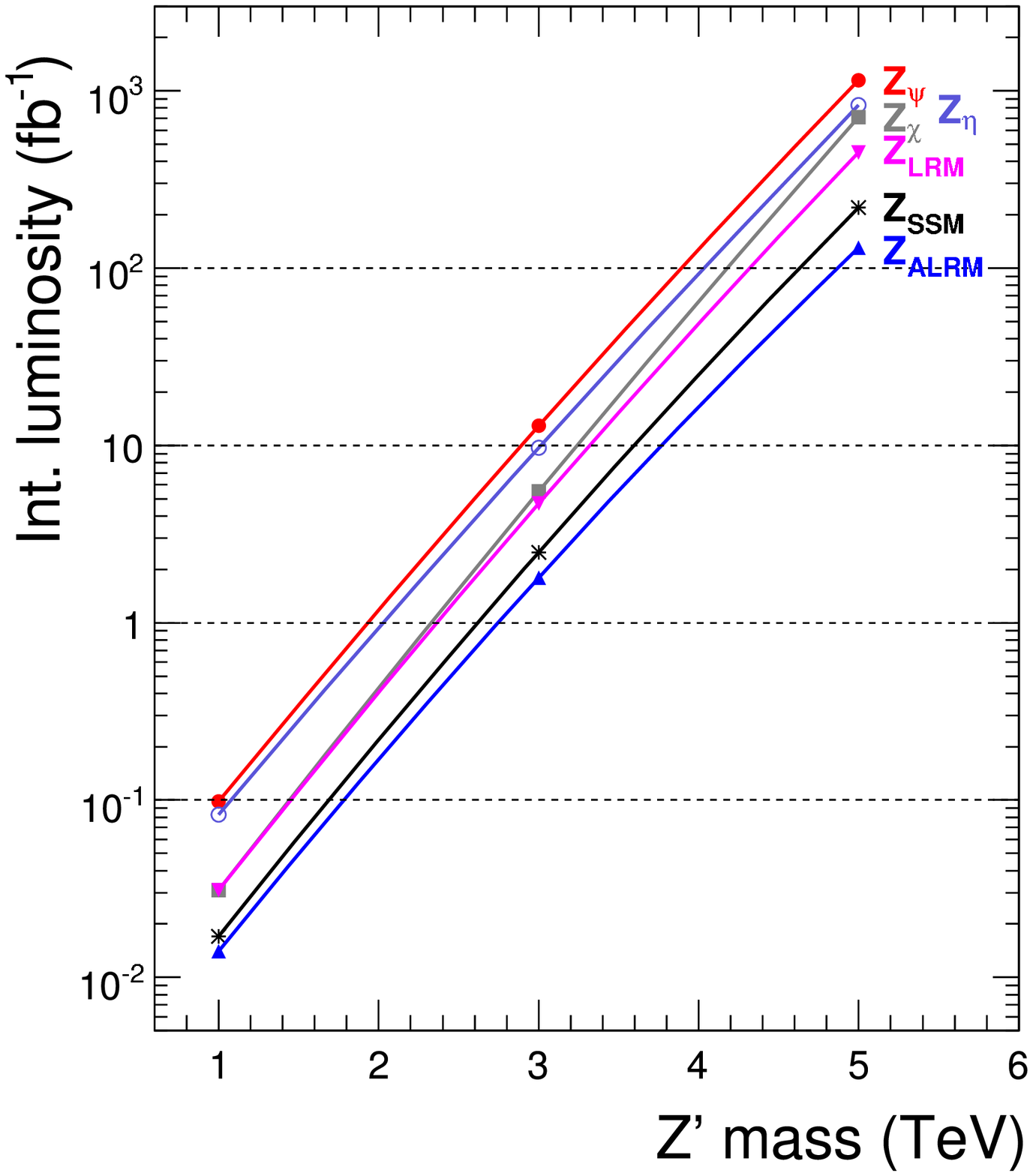,height=4.0cm} \hspace*{2mm}
    \psfig{figure=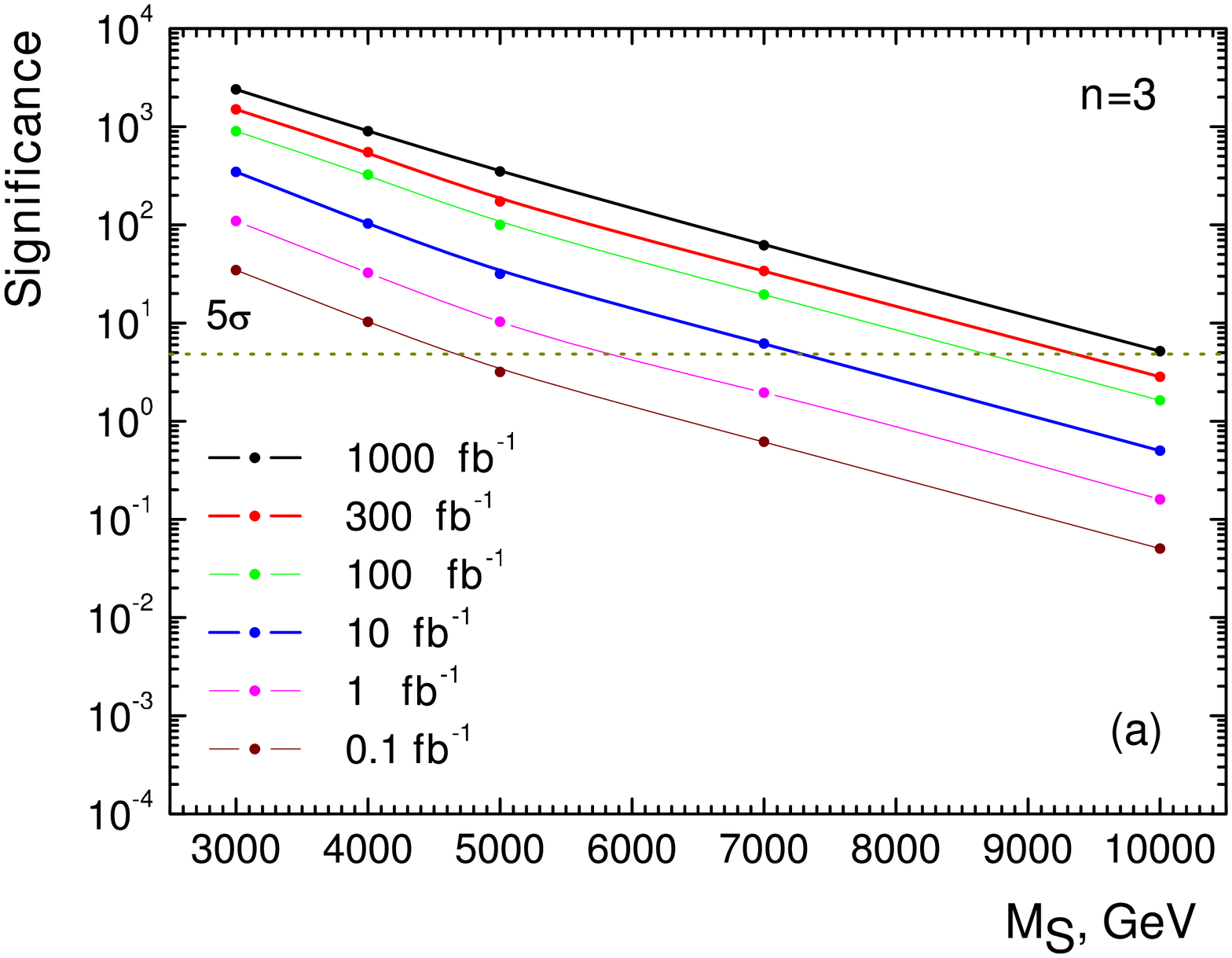,height=4.0cm} \hspace*{2mm}
    \psfig{figure=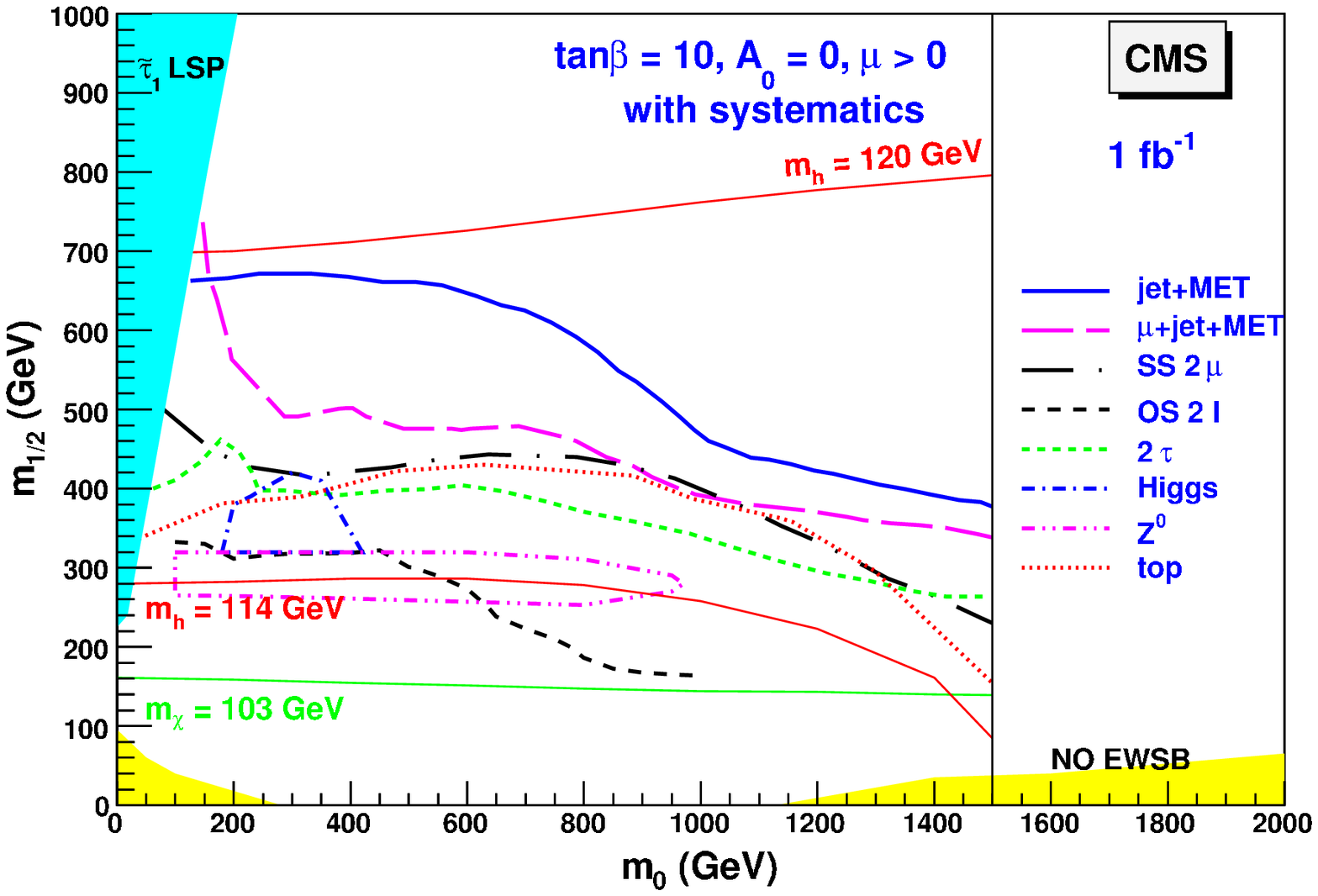,height=4.0cm}
  \end{center}
  \caption{Left: integrated luminosity needed for a Z' discovery in CMS,
  as a function of Z' mass, in various models with different Z' couplings.
  Center: significance of a discovery of large extra dimensions in the ADD
  model for various luminosities, for three extra dimensions,
  as a function of the fundamental Planck
  scale $M_S$. Right: CMS reach in the mSUGRA parameters $m_{1/2}$ and $m_0$,
  for $\tan \beta = 10$, $A = 0$, $\mu > 0$, for 1 fb$^{-1}$ of data.
  \label{fig:zprimereach}}
\end{figure}

As shown in Figure~\ref{fig:zprimereach}, a Z' could be discovered with 1 fb$^{-1}$ up
to masses of 2-2.8 TeV, depending on the Z' couplings, Randall-Sundrum
graviton resonances up to 2.3 TeV, and ATLAS and CMS would be sensitive
to a 6 TeV fundamental Planck scale if there would exist three large 
extra dimensions of the type of the ADD model (4 TeV for six extra dimensions).

%%%%%%%%%%%%%%%%%%%%%%%%%%%%%%%%%%%%%%%%%%%%%%%%%%%%%%%%%%%%%%%%%%%%%%%%%%%%%%%
%
\section{Higgs search in the WW $\rightarrow \ell \ell \nu \nu$ channel}

One of the most promising Higgs search channels with early data is the
search for a Higgs in the decay channel H $\rightarrow$ WW
$\rightarrow \ell \ell \nu \nu$, for a Higgs boson in the mass
range 150-170 GeV. CMS estimates that a discovery can be made with less
than 1 fb$^{-1}$ of data~\cite{droz}, and ATLAS also considers this, in
particular for Higgs production through WW fusion, the most promising
channel.

Since there are two neutrinos in the final state,
the mass resolution is poor, and this is essentially a counting experiment.
Therefore, it is extremely important to have a good understanding of
the background. The major backgrounds, WW and top-quark pair production,
are extracted from the data itself through a procedure involving several
control samples. The uncertainties related to this procedure have been
studied in detail, and seem to be under control. A small, but important,
component to WW production comes from gluon-gluon fusion processes.

%%%%%%%%%%%%%%%%%%%%%%%%%%%%%%%%%%%%%%%%%%%%%%%%%%%%%%%%%%%%%%%%%%%%%%%%%%%%%%%
%
\section{Supersymmetry}

Supersymmetry (SUSY) is a theoretically attractive candidate for physics beyond
the Standard Model, and there are several arguments in favor of supersymmetry
at the TeV scale, accessible at the LHC. If the LHC does not find
any evidence for supersymmetry, it is extremely unlikely that supersymmetry
is the answer to open issues in the Standard Model like the hierarchy problem.
Also the interpretation of the lightest supersymmetric particle as a dark
matter candidate will be problematic, certainly if direct detection and
astroparticle physics experiments also do not find evidence.

\begin{figure}[htbp]
  \begin{center}
    \psfig{figure=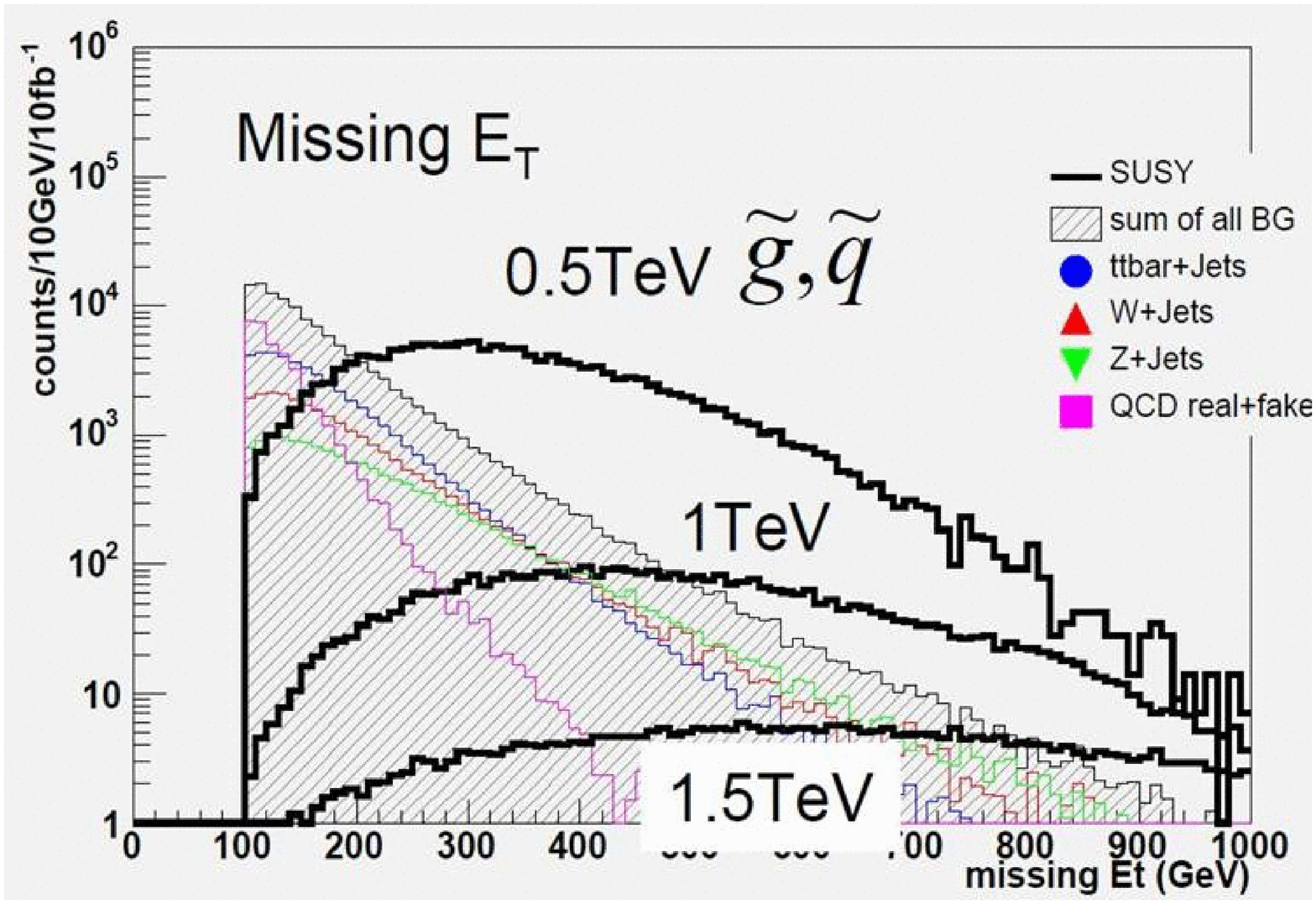,height=3.8cm} \hspace*{1mm}
    \psfig{figure=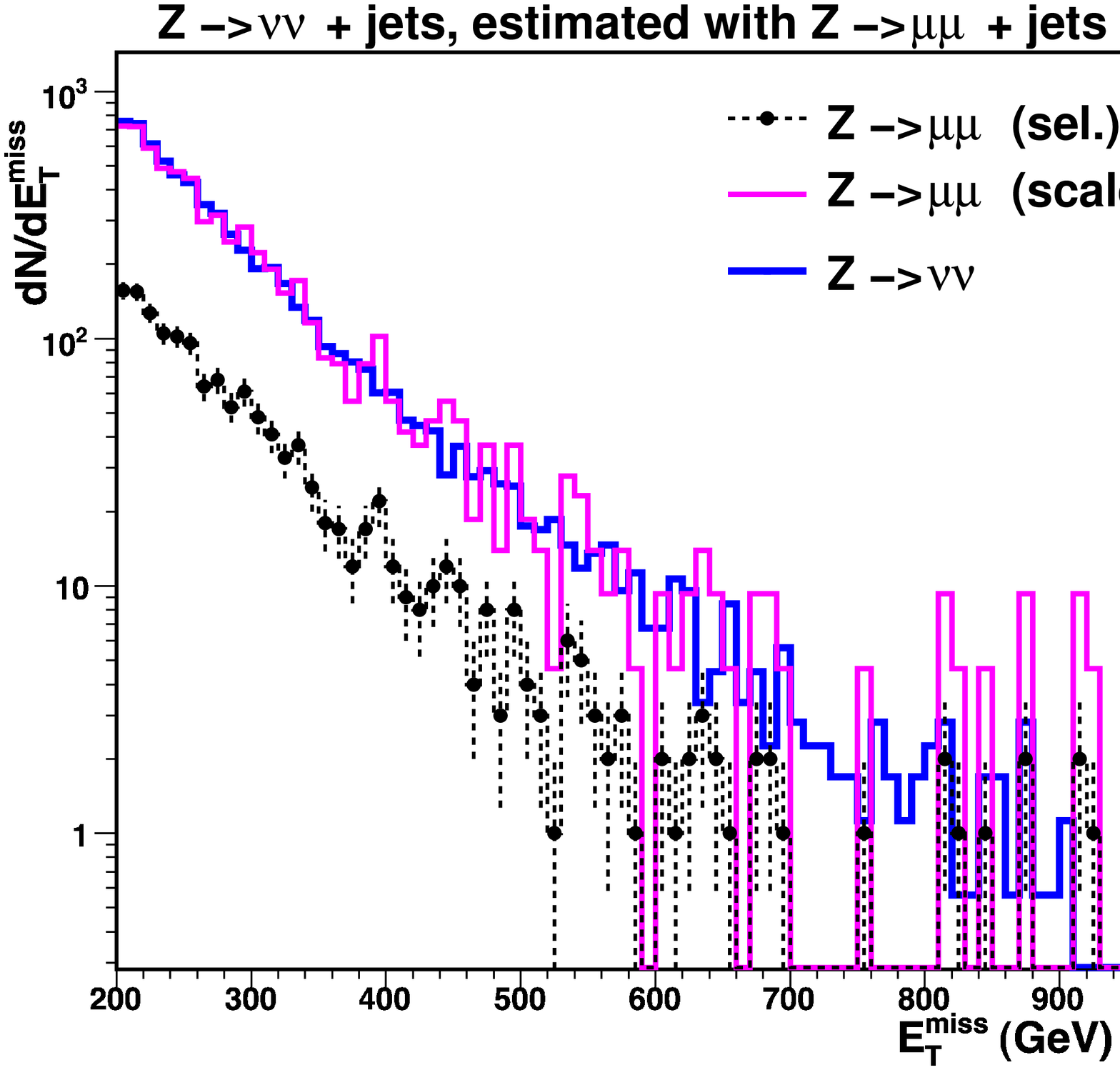,height=3.8cm} \hspace*{1mm}
    \psfig{figure=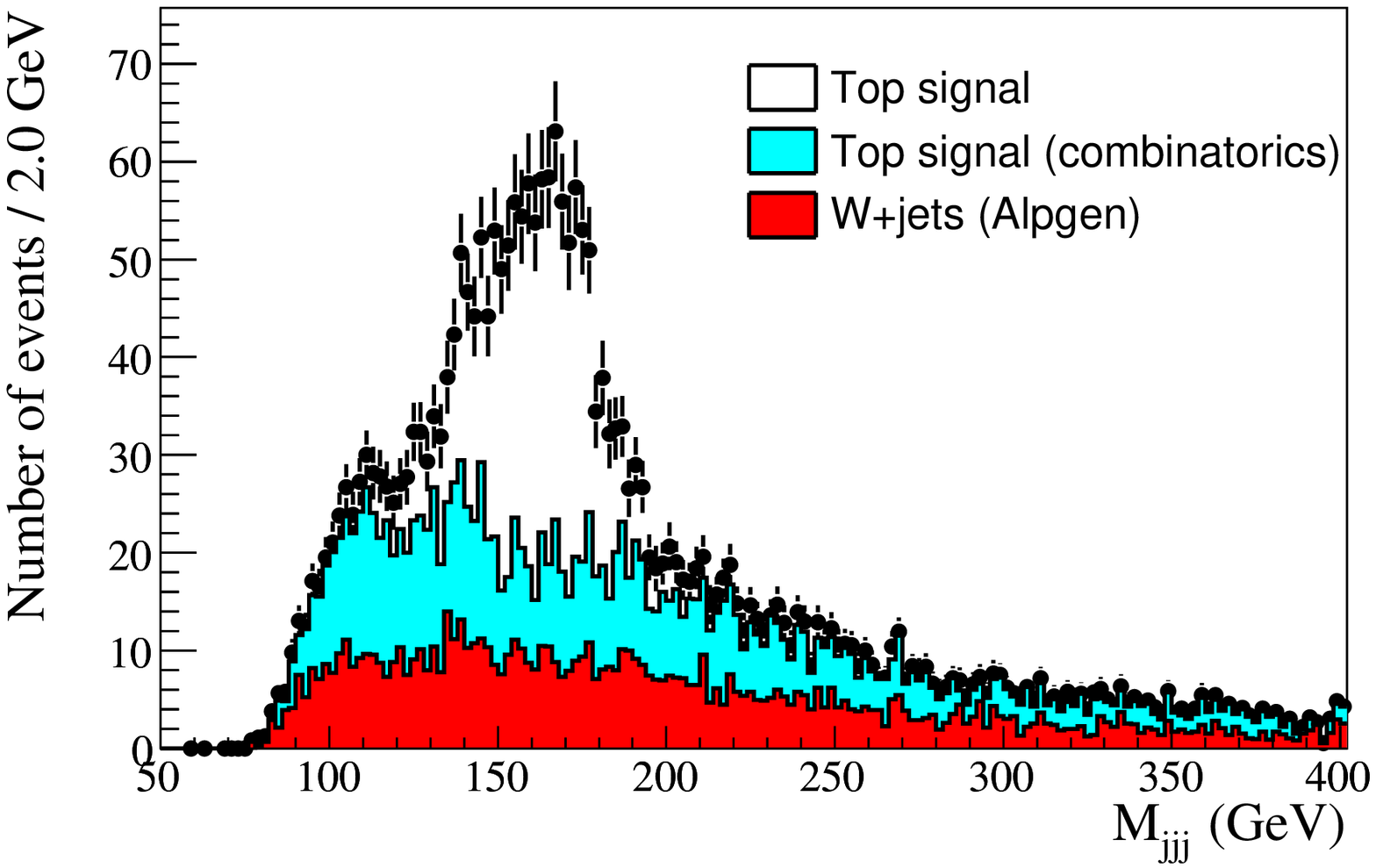,height=3.8cm,width=5.2cm}
  \end{center}
  \caption{Left: \met $\,$ distribution in multi-jets final states,
           for potential SUSY signals
           at 0.5, 1.0 and 1.5 TeV mass scales, and \met $\,$
           in various backgrounds.
  Center: \met $\,$ distribution of Z ($\rightarrow \nu \nu$) plus
  at least two jets events, and how this background to SUSY could be estimated from
  Z ($\rightarrow \mu \mu$) events, properly scaled.
  Right: Expected signal for top-quark pairs in 100 pb$^{-1}$ of early
   ATLAS data, without using b-tagging. Shown is the invariant mass
   distribution of the three jets assigned to a hadronically decaying top quark.
  \label{fig:etmisssusy}}
\end{figure}

If the SUSY mass scale is only just above the Tevatron limits, SUSY is the
prime candidate for an early discovery at the LHC. A search for
SUSY in early data must be robust (able to cope with background
uncertainties and a non-optimal detector) and general, yet efficient.
Excellent opportunities exist in final states with high $E_T$ jets, and 
significant missing transverse energy (\met).
In order to suppress the QCD background and facilitate triggering, 
it is possible to further
demand one (or more) high $p_T$ lepton(s). Such final states typically arise
in the decay chains of squarks and gluinos, which will be copiously produced
at the LHC, if kinematically allowed. Assuming R-parity conservation, the lightest
SUSY particle will escape the detector unseen; if R-parity is violated there
will be little \met , but still events with many high $E_T$ jets.
Standard Model backgrounds to the SUSY search mainly come from top quark pair 
production, Z or W boson production in association with jets, and QCD jet 
production. An example of the \met $\,$ distribution in signal
and background is shown in Figure~\ref{fig:etmisssusy} (left).

There are several challenges for an early SUSY discovery: reconstruction of
leptons, jets, and \met $\,$ in busy events; fake \met $\,$ from detector effects;
trying to be general, yet efficient, in the SUSY search;
and understanding the Standard Model background well.

The \met $\,$ is a measure of energy carried away by escaping, unmeasured 
particles, such as neutrinos, or (semi)stable weakly interacting particles
in new physics models.  Measuring \met $\,$ relies on accurate reconstruction 
of the transverse momentum balance, and is affected by detector effects 
(holes, noise, punchthrough), and a finite resolution. QCD jets can have 
real \met $\,$ due to neutrinos, and it is a challenge to understand
the high \met $\,$ tail well.

Reconstruction of objects in a busy environment can be tested at the LHC in
top-quark pair production events, which constitute an excellent calibration
sample, and also provide interesting physics. At the LHC, the ratio of
production of top quark pairs and production of background to top quark
pairs is more favorable than at the Tevatron, and clean samples of $t \bar{t}$
events can be selected, in particular in events with at least one electron
or muon. ATLAS, for example, expects a signal as shown in 
Figure~\ref{fig:etmisssusy}(right) in 100 pb$^{-1}$ of early data, even
without b-tagging. Such a sample will be useful for the jet energy
scale calibration, and calibration of b-tagging and \met. With a working b-tagging,
very clean $t \bar{t}$ samples can be selected.

% \begin{figure}[htbp]
  % \begin{center}
    % \psfig{figure=topsignal.eps,height=1.5in}
  % \end{center}
  % \caption{Expected signal for top-quark pairs in 100 pb$^{-1}$ of early
   % ATLAS data, without using b-tagging. Shown is the invariant mass
   % distribution of the three jets assigned to a hadronically decaying top quark.
  % \label{fig:topsignal}}
% \end{figure}

The estimation of backgrounds in SUSY searches is performed as much as 
possible from data. 
% The main sources are top-quark pair production, and W 
% and Z production in association with extra jets. 
As an example
(Figure~\ref{fig:etmisssusy}(center)) the important Z ($\rightarrow \nu \nu$) plus
jets background in the SUSY jets plus \met $\,$ channel can be calibrated with 
a clean Z ($\rightarrow \mu \mu$) plus jets sample, but also with 
W ($\rightarrow \mu \nu$) plus jets samples. Many other control samples
are under study.

% \begin{figure}[htbp]
%   \begin{center}
%     \psfig{figure=znunu.eps,height=1.5in} \hspace*{1cm}
%     \psfig{figure=cms1fb.eps,height=1.5in}
%   \end{center}
%   \caption{Missing $E_T$ distribution of Z ($\rightarrow \nu \nu$) plus
%   at least two jets, and how this background could be estimated from
%   Z ($\rightarrow \mu \mu$) events, properly scaled.
%   CMS reach in the mSUGRA parameters $m_{1/2}$ and $m_0$,
%            for $\tan \beta = 10$, $A = 0$, $\mu > 0$, for 1 fb$^{-1}$ of data.
%   \label{fig:znunu}}
% \end{figure}

% \begin{figure}[htbp]
  % \begin{center}
    % \psfig{figure=cms1fb.eps,height=1.5in}
  % \end{center}
  % \caption{CMS reach in the mSUGRA parameters $m_{1/2}$ and $m_0$,
           % for $\tan \beta = 10$, $A = 0$, $\mu > 0$, for 1 fb$^{-1}$ of data.
  % \label{fig:susyreach}}
% \end{figure}

%%%%%%%%%%%%%%%%%%%%%%%%%%%%%%%%%%%%%%%%%%%%%%%%%%%%%%%%%%%%%%%%%%%%%%%%%%%%%%%
%
\section{Final comments}

The first LHC data is eagerly awaited by a large community of experimentalists
and theorists. There will be a strong pressure on the experiments to
provide results early, and it is hard to prevent high-profile analyses to
take place in a ``glass box''. There will also be strong internal competition
within ATLAS and CMS, and between ATLAS and CMS. It is important not to
compromise on the quality of the results, after all the LHC will be operating
in new territory.
In this sense, the issue of blind analyses at the LHC, in order to
prevent any bias, has come up, but has
also raised internal discussions (at least in ATLAS) on the feasibility.

% Just from statistical fluctuations, we can expect a certain number of
% excesses over background even in the absence of new physics: after all more
% than 1000 physicists will probably examine at least 100 histograms with 100 bins
% each. The experiments must of course perform their usual checks and come out
% with the full story, but it is also important that they decide beforehand what to
% do. In this sense, the issue of blind analyses at the LHC has come up, but has
% also raised internal discussions (at least in ATLAS) on the feasibility.

ATLAS and CMS are learning more and more from CDF and D0, not the least since
many CDF and D0 physicists are joining ATLAS and CMS. There is still a lot
to be learned from the Tevatron in terms of analysis techniques and background
estimations from data, and on W and Z production. Certainly, CDF and D0 have shown
that understanding the detector with data will be a major challenge.

%%%%%%%%%%%%%%%%%%%%%%%%%%%%%%%%%%%%%%%%%%%%%%%%%%%%%%%%%%%%%%%%%%%%%%%%%%%%%%%
%

\end{document}